# Monte Carlo and MD Modeling of Density Relaxation by Tapping

Oleksandr Dybenko, Anthony D. Rosato, David J. Horntrop*, Vishagan Ratnaswamy, and Lou Kondic*

*Granular Science Laboratory, Department of Mechanical Engineering*
*\*Department of Mathematical Sciences and Center for Applied Mathematics and Statistics*
*New Jersey Institute of Technology, Newark, NJ 07102*

The density relaxation phenomenon is modeled using both Monte Carlo and dissipative MD simulations to investigate the effects of regular taps applied to a vessel having a planar floor filled with monodisperse spheres. Results suggest the existence of a critical tap intensity that depends on the mass overburden, which produces a maximum bulk solids fraction. We find that the mechanism responsible for the relaxation phenomenon is evolving quasi-ordered packing structure propagating upwards from the plane floor.

Density relaxation describes the phenomenon in which granular solids undergo an increase in bulk density as a result of properly applied external loads. The ability of granular materials to undergo density changes is an inherent property that is not well-understood, and thus it remains a critical impediment in developing predictive models of flowing bulk solids. From a historical perspective, one may consider the phenomenon as having its basis in the rather extensive literature on the packing of particles [1], where the concern was often in characterizing loose and dense random structures.

The focus of early studies on density relaxation was on how optimal packings could be produced through the use of continuous vibrations or discrete taps [2]. Recently, there has been a resurgence of interest (partly triggered by experiments [3]) in uncovering particle-level mechanisms responsible for observed macroscopic behavior. These experimental results have spurred theoretical studies involving free volume arguments [4], parking lot paradigms [5], stroboscopic decay approaches [6], and mesoscopic one-dimensional lattice models [7]. Computational investigations involving stochastic and deterministic simulations [8-10] have also been carried out to directly address the coupling between the detailed particle dynamics and bulk behavior. In this context, we note the single particle and collective dynamics observed in [8], filling mechanisms [11], and the appearance of local structural order evidenced by a second peak in the radial distribution function [12]. Important progress related to density relaxation has also been made in characterizing glass and jamming transitions in granular materials [13].

The results reported in this letter provide significant new insight regarding the process of density relaxation in granular materials exposed to discrete taps imposed by the motion of a flat plane floor. The process is modeled through (1) hard-sphere Monte Carlo (MC) simulations, and (2) dissipative molecular dynamics (DMD) using inelastic, frictional spheres. Although these particle-based methods are quite different (one purely stochastic and the other deterministic), both reveal a clear picture of the dynamical process responsible for density relaxation, namely, the upward progression of self-organized layers induced by the plane floor as the taps evolve. Indeed, its occurrence in both the MC and DMD simulations suggests the universality of this mechanism in density relaxation which, to our knowledge, has not been previously reported in the literature. We also unambiguously demonstrate the importance of the nature of the applied tap in the temporal evolution of these granular systems.

The remainder of this letter is organized as follows. We first describe our MC simulation model (reported in detail elsewhere), which is based on a modification of the standard Metropolis scheme. In this approach, the effect of taps applied to a containment vessel filled with spherical particles of diameter $d$ is idealized by applying a normalized vertical separation $\Delta/d \equiv \gamma$ of the particle assembly from the floor. We demonstrate the importance of incorporating this separation (as compared to employing a uniform spatial expansion) to dislodge metastable configurations in attaining an increase in density. We find evidence of a critical tap intensity that optimizes the evolution of packing density. Finally, we discuss our DMD simulation model and results revealing the same tap-induced ordering effect of the floor on the local microstructure.

The computational domain is a box having a solid plane floor, and periodic boundary conditions in the lateral directions ($x$ and $z$; $12d$ x $12d$). Spheres that are at the outset randomly placed in the box, settle under gravity to a loose poured assembly. (See Fig. 1a). The poured assembly is then subjected to a series of taps of intensity $\gamma$.

In our MC simulation, a single step in the algorithm consists of the random selection of a particle

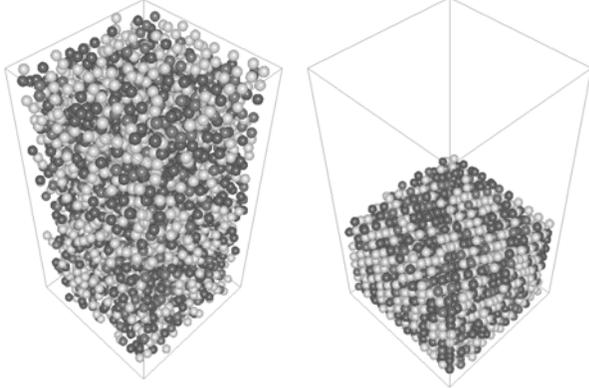

**Fig. 1.** Snapshot of a system of hard spheres within the periodic box for: (a) an initial, random state and (b) a configuration of the ensemble tapped at $\gamma = 0.25$ for which $\langle v \rangle = 0.6542$.

$\mathbf{x}_i = (x_i, y_i, z_i)$, followed by its assignment to a trial position $\mathbf{x}'_i = \mathbf{x}_i + \delta(1-2\xi)$, where $\xi$ is a random vector sampled from a uniform distribution on $(0,1)$, and $\delta$ is a maximum allowed displacement. The trial position is accepted unconditionally (provided that an overlap does not occur) as the new location if the change in the system energy $\Delta E \equiv mg \sum_{i=1}^{N}(y'_i - y_i) < 0$. Otherwise, ($\Delta E \geq 0$) the trial position is accepted with probability $e^{-\beta \Delta E}$. For the macroscopic particles that we are studying, $\beta$ is very large ($\beta \Delta E \gg 1$) so that the likelihood of an accepted upward displacement is small. Another particle is then selected at random, and the above procedure is repeated. As this settling process advances through many thousands of MC steps, the mean free path decreases resulting in a drastically slowed down rate of approach to a local equilibrium. Therefore, the parameter $\delta$ is modified every $10^4$ MC steps in accordance to $\delta' = 0.995\delta$ if fewer than half of these steps are accepted.

An individual tap consists of a separation $\gamma$ followed by the settling process as described above. Within a single tap $n$ of intensity $\gamma$, we monitored the bulk solids fraction $v_j(n;\gamma)$ (i.e., the fraction of volume occupied by spheres) at MC step $j$ every $10^6$ steps. The settling process was terminated when

$$\left| v_{(k+1)10^6}(n;\gamma) - v_{10^6 k}(n;\gamma) \right| < 0.001, \ k = 0, 1, 2, \ldots.$$

This protocol was validated by enforcing the latter criterion twice to ensure that premature termination could not occur.

In all cases reported in this letter, poured assemblies with solids fractions $v_o$ in the range 0.56 to 0.58 filled the periodic box to a depth of approximately $22d$. For each tap $n$, we computed the ensemble averaged bulk solids fraction $\langle v(n;\gamma) \rangle$ over $M$ realizations, and its standard deviation

$$\langle \sigma(n;\gamma) \rangle = \sqrt{\frac{1}{M} \sum_{k=1}^{M} \left[ v_k(n;\gamma) - \langle v(n;\gamma) \rangle \right]^2}$$ where

$\langle v(n;\gamma) \rangle = \frac{1}{M} \sum_{k=1}^{M} v_k(n;\gamma)$. The equilibrium bulk solids fraction $v_\infty(\gamma)$ was found by increasing the number of realizations over a sufficient number of taps $N_T$, until the condition $\langle \sigma(N_T;\gamma) \rangle < 0.001$ was satisfied.

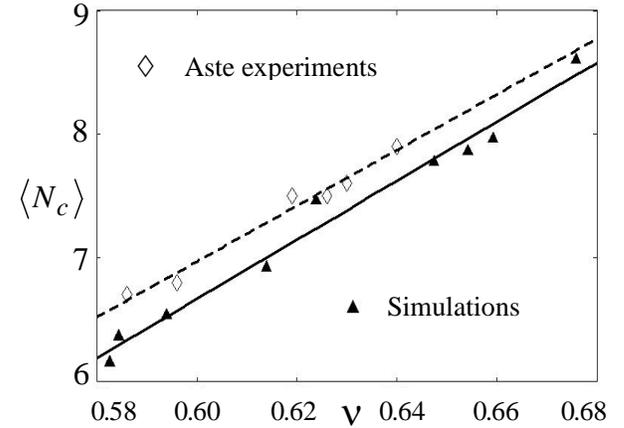

**Fig. 2.** Mean coordination number $\langle N_c \rangle$ versus solids fraction $v$ from the MC simulation (▲) and experimental measurements (◊) of Aste et al. [11]. The straight lines are linear regression fits to show the trends of experimental (dashed) and computational (solid) results.

The simulated mean coordination number in Fig. 2 (i.e., average number of contacts per particle) versus solids fraction was in reasonably good agreement with the experimental measurements of Aste et al. [12], showing the same trend over a range of $v$. Since we do not know the error of the experimental results, we cannot judge the significance of the slight shift of these two lines. Although not shown here, at low tap intensities, the solids fraction evolution data fit the inverse log phenomenological law [3].

Our MC tapping procedure differs from that used in [8], where vertical position-dependent displacements of the particles ($y' = \lambda y$) were applied in sync with random lateral perturbations. For the packings (~22d high) reported in this letter, application of that tapping mode produced little or no increase in solids fraction in the upper portion of the bed, so that the bulk solids fraction remained relatively low. We attribute this behavior to an overly aggressive vertical displacement of the particles with distance from the floor, so that after each tap, the system tends to lose memory of its previous microstructure. Physically, this corresponds to the situation in which vigorous taps or shakes are applied so that the system relaxes to nearly the same or lower bulk density. Although exactly the same potential energy was introduced for both types of tapping, Fig. 3 shows that a significantly higher densification rate occurred when the entire particle assembly was raised from the floor, thus demonstrating the importance of the choice of model for a tap to obtain rapid equilibration of the system.

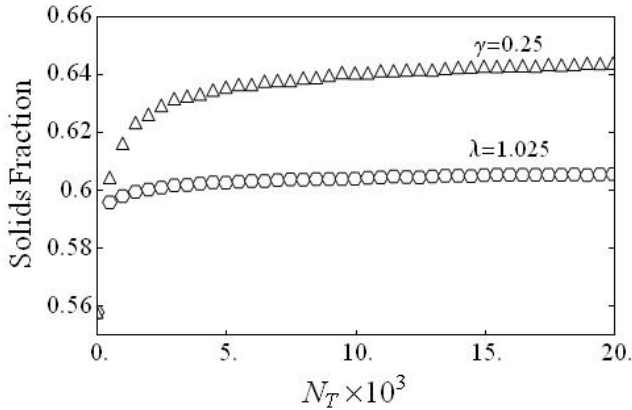

**Fig. 3.** Solids fraction versus tap number for two tap modes: $\gamma = 0.25$ and $\lambda = 1.025$, each producing the same increase in system potential energy.

A series of case studies for $0.1 \leq \gamma \leq 0.5$ was carried out, from which the corresponding bulk solids fractions were determined. Our findings (summarized in Fig. 4.) revealed a strong dependency of the densification rate on $\gamma$, for which a rapid ($N_T \sim 10^3$) equilibration to $v_\infty(\gamma)$ was achieved for $\gamma \geq 0.3$. Although we did not run to equilibrium for smaller $\gamma$ because of the large number of taps required, the inset suggests the existence of a critical intensity yielding an optimal packing. This finding is also supported by our experiments reported elsewhere [14], and by the early conjecture in [15] that bulk density is related to impact velocity.

Fig. 5 illustrates the mechanism responsible for density relaxation in our Monte Carlo model – namely, the upward advance of self-organized layers induced by the plane floor. The progression is quantified by the ensemble-averaged fraction of the total number of sphere centers $\bar{n}$ as function of the distance $y/d$ from the floor. For the untapped poured system, $\bar{n}$ (Fig. 5a) is uniform except for a well-known ordering [16] of the first few layers adjacent to the floor. As the tapping proceeds, there is an upward advance of the layering: after 50 and 250 taps (Figs. 5b,c), peaks in $\bar{n}$ have formed near the floor, while at 90,000 taps (Fig. 5d), the peaks appear throughout the depth.

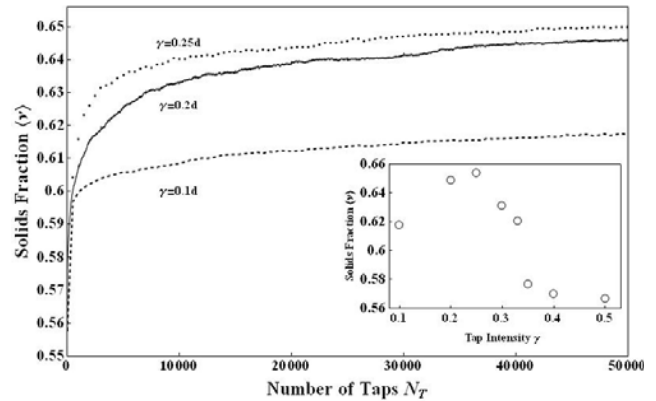

**Fig. 4.** Bulk solids fraction $\langle v \rangle$ versus the number of taps $N_T$ at $\gamma = 0.1, 0.2, 0.25$. The inset shows $\langle v \rangle$ for $0.1 \leq \gamma \leq 0.5$.

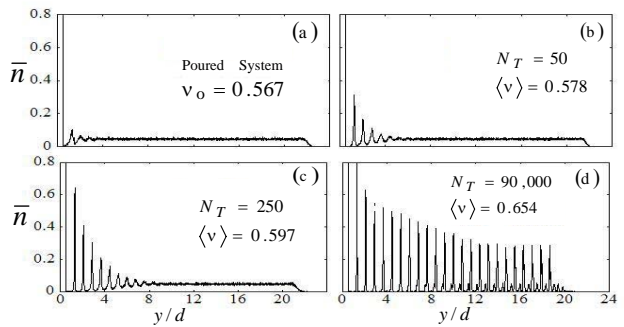

**Fig. 5.** MC ensemble-averaged fraction of the total number of sphere centers $\bar{n}$ as a function of $y/d$ for tap intensity $\gamma = 0.25$: (a) poured system, $v_0 = 0.567$; (b) 50 taps, $<v> = 0.578$; (c) 250 taps, $<v> = 0.597$; (d) 90,000 taps, $<v> = 0.654$.

A growth in the magnitude of the peaks as the taps evolve is a manifestation of the gradual filling of voids. The locations of the first four peaks correspond exactly to a hexagonal close-packed structure. The increase from $v_0 = 0.567$ to $<v> = 0.654$ is accompanied by an (approximate) $2d$ reduction of the bed's effective height from its initial level. In Fig. 5d one can also see smaller secondary peaks in the upper side of the bed, showing existence of (yet) an imperfect structure there. These peaks gradually vanish as the tapping progresses.

In order to ensure that the ordering observed in the MC model is a universal mechanism, we carried out an analogous study via DMD. Particles were inelastic, frictional spheres obeying binary, soft-sphere interactions [17] in which normal and tangential impulses are functions of a small overlap between colliding particles. Energy loss in the normal direction (i.e., along center line of contacting spheres) is produced by linear loading ($K_1$) and unloading springs ($K_2$), corresponding to a constant restitution coefficient $e = \sqrt{K_1/K_2}$. This model has been shown to reproduce the nearly linear loading behavior for a spherical surface that experiences plastic deformation of the order of 1% of a particle diameter. In the tangential direction, a Mindlin-like model [18] is used in which tangential stiffness decreases with tangential displacement until full sliding at the limit μ occurs.

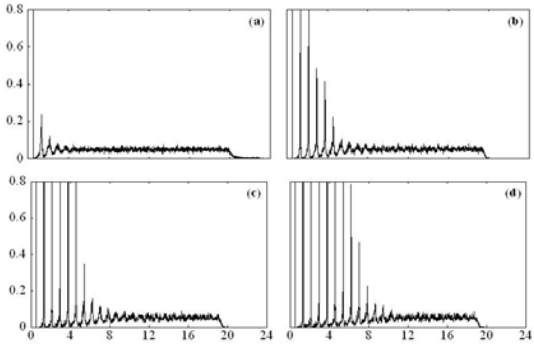

**Fig. 6**: DMD computations of $\bar{n}$ for $\tau = 0.0667\,\text{s}$, $t_r = 0.3\,\text{s}$, $a/d = 0.35$ ($\Gamma = a\omega^2/g = 2.03$). Each plot is an average over 10 taps, and is labeled by the end time over which the average was computed.

The computational domain is again a laterally periodic box ($12d \times 12d$) in which particles settle under gravity to a fill height of approximately $21d$ at $v_o = 0.5677$. Subsequently, the granular bed is tapped by applying harmonic oscillations to the floor consisting of a half-sine displacement (amplitude $a$ and period $\tau$), followed by a relaxation time $t_r$ of sufficient duration to ensure that upon collapse, a quiescent state of minimal kinetic energy is attained. $\bar{n}$ is averaged over 10 consecutive relaxed configurations. Fig. 6 illustrates the results for the case $\tau = 0.0667\,\text{s}$, $t_r = 0.3\,\text{s}$, $a/d = 0.35$. The material density of the spheres corresponded to acrylic (ρ = 1200 kg/m$^3$), and we chose $e = 0.9$ and μ = 0.1. At $t = 366.7$s (or 1000 taps), a bulk solids fraction of $\langle v \rangle = 0.657$ is attained. The qualitative behavior of $\bar{n}$ is analogous to the MC results (Fig. 5) in that peaks grow and advance from the floor as the taps evolve. In future work, we will carry out a broad systematic DMD study of dilation behavior and its relationship to the relaxed state, including the dependence of the relaxation process on material parameters.

In this Letter, the results of our MC and DMD investigations of tapped density relaxation were reported. Good agreement of the MC-generated coordination number versus solids fraction with reported experiments was found, along with strong correlation with an inverse-log phenomenology. We find compelling evidence of a critical tap intensity that optimizes the evolution of packing density. Both our stochastic (MC) and deterministic (DMD) models revealed the same dynamical process responsible for density relaxation, namely, the progression of self-organized layers induced by the plane floor as the taps evolve. We expect that this prediction will be of interest to numerous scientific and engineering problems involving self-organization and its relation to approach to equilibrium.

The authors thank D. Blackmore and P. Singh for discussions of this work. LK and DJH acknowledge partial support from NSF-DMS-0605857 and NSF-DMS-0406633, respectively. Partial computational support was provided by NSF-DMS-0420590.